Letter

# Linear dichroic soft x-ray microscopy of ferroelectric stripe domains in epitaxial $K_{0.6}Na_{0.4}NbO_3$

M. Schneider,[1] T. A. Butcher,[1,*] S. Wagner,[1] D. Metternich,[1,2,3] C. Klose,[1] E. Malm,[4] R. Battistelli,[2,3] V. Deinhart,[1] J. Fuchs,[1] S. Wittrock,[2] T. Karaman,[2,3] K. Puzhekadavil Joy,[2,3] M. Patra,[2,3] F. Büttner,[2,3] S. Wintz,[2] M. Weigand,[2] C. M. Günther,[5] D. Engel,[1] P. Gaal,[6,7] J. Schwarzkopf,[6] B. Pfau,[1] and S. Eisebitt[1,8]

[1]Max Born Institute for Nonlinear Optics and Short Pulse Spectroscopy, 12489 Berlin, Germany
[2]Helmholtz-Zentrum Berlin für Materialien und Energie, 14109 Berlin, Germany
[3]Experimental Physics V, Center for Electronic Correlations and Magnetism, University of Augsburg, 86159 Augsburg, Germany
[4]MAX IV, Lund University, 224 84 Lund, Sweden
[5]Technische Universität Berlin, Zentraleinrichtung Elektronenmikroskopie, 10623 Berlin, Germany
[6]Leibniz-Institut für Kristallzüchtung, Max-Born-Straße 2, 12489 Berlin, Germany
[7]TXproducts UG, Luruper Hauptstrasse 1, 22547 Hamburg, Germany
[8]Technische Universität Berlin, Institut für Physik und Astronomie, Straße des 17. Juni 135, 10623 Berlin, Germany

(Received 28 March 2026; revised 4 September 2026; accepted 8 September 2026; published 2 October 2026)

Functional properties of ferroelectric thin films are governed by domains that can be engineered by epitaxial strain. Soft x-ray microscopy can image domain structures with elemental and electronic sensitivity, but hitherto its application to strain-stabilized domains has been hindered by the absorption of soft x-rays in epitaxial substrates. Here, it is demonstrated how this limitation can be overcome by locally back-thinning the (110) $TbScO_3$ substrate of epitaxial $K_{0.6}Na_{0.4}NbO_3$ ferroelectric thin films to achieve soft x-ray transparency at the O *K*-edge around 530 eV. Strain-induced ferroelectric stripe domains with periods down to 44 nm were resolved by scanning transmission x-ray microscopy and coherent diffractive imaging by exploiting the x-ray linear dichroism of hybridized O 2*p*–Nb 4*d* states, providing sensitivity to in-plane polarization components under normal incidence. The results establish soft x-ray microscopy for nanoscale imaging of epitaxial ferroelectric domain structures and open perspectives for time-resolved studies thereof.

Functional properties of ferroelectric materials such as piezoelectricity, flexoelectricity, and photoelectric performance depend on the arrangement, orientation, and size of the ferroelectric domains and their domain walls [1,2]. Domain formation is governed by mechanical and electrical boundary conditions of the films, which can be engineered in epitaxial thin films by using lattice mismatched substrates and the application of electrode layers, respectively. As the size of ferroelectric domains in thin films is on the nanoscale, microscopy methods with high spatial resolutions are required for their study. The common approach is piezoresponse force microscopy (PFM) with usual spatial resolutions around 20 nm [3]. Alternative and complementary characterization techniques with high spatial resolution are x-ray microscopy on the nanoscale or transmission electron microscopy (TEM) on the unit-cell level. The latter is impractical for extended regions of ferroelectric domains in standard setups [4].

The availability of high coherent flux and tunable photon energies at synchrotron light sources has enabled imaging of ferroelectric order both in the hard and soft x-ray energy regimes. High-resolution x-ray microscopy can be separated into three approaches: Firstly, full-field techniques that provide direct images such as x-ray photoemission electron microscopy (X-PEEM) [5] or dark field x-ray microscopy [6,7]. Secondly, scanning microscopy relying on focusing the x-rays to a sub-100-nm diameter spot, which is usually achieved by Fresnel zone plates (FZPs). Thirdly, methods based on the reconstruction of real-space images from coherent diffraction patterns by coherent diffractive imaging (CDI) techniques. Regarding hard x-rays, information about ferroelectric domain structures is encoded in alterations of Bragg peaks. This structural information can be related to ferroelectric domains by mapping the Bragg peaks in nanoprobe measurements with FZP focused hard x-ray beams in reflection geometry [8–10]. Less well established is the imaging of ferroelectric domain structures with the CDI approach of Bragg ptychography, which has been employed for imaging out-of-plane ferroelectric polarization with sub-10-nm spatial resolution [11].

Soft x-ray microscopy relies on x-ray beams with lower photon energies between 200–2000 eV and a higher coherent fraction than hard x-rays. Contrast mechanisms arise due to electronic transitions at elemental x-ray absorption edges contained in this energy range. As ferroelectrics are predominately oxides, the O *K*-edge around 530 eV with the transition of $1s \rightarrow 2p$ is of high relevance for their study. This transition provides strong sensitivities to anisotropic charge distributions in the unit cell as well as the hybridization of O 2*p* orbitals with metal states, which can yield the ferroelectric domain structures by x-ray linear dichroism (XLD). Linear dichroic imaging senses the axis of the ferroelectric polarization relative to the plane of polarization of the x-ray beam. The direction in which the ferroelectric polarization vector points cannot be recovered from such measurements, unlike in the case of imaging with the x-ray magnetic circular dichroism (XMCD) contrast for ferromagnetic materials.

*Contact author: tim.butcher@mbi-berlin.de

At the O *K*-edge, XLD imaging was first carried out with surface-sensitive X-PEEM [5] and recently by the CDI method of soft x-ray ptychography [12] in multiferroic $BiFeO_3$ (BFO) [13]. Strong absorption of soft x-rays at the elemental absorption edges restricts the sample thicknesses to sub-200 nm for high-resolution soft x-ray microscopy that is performed in transmission, which has limited linear dichroic imaging to membranes of ferroic oxides [13–17]. This limitation has so far prevented the application of linear dichroic soft x-ray microscopy to epitaxial thin films in their as-grown state on substrates. Although it is possible to perform ptychography in reflection geometry [18], this involves a loss in spatial resolution due to the suppressed scattering cross section compared to transmission measurements.

Previous studies with soft x-ray ptychography were carried out on freestanding BFO films [13,16,17], which invariably form ferroelectric mosaic domains due to the lack of epitaxial strain. However, the stabilization of ferroelectric phases in new materials and useful domain structures relies on epitaxial strain from a substrate. Imaging such epitaxially stabilized domain structures by soft x-ray microscopy therefore requires a transmission geometry compatible with thick oxide substrates. The present study addresses this challenge.

An ongoing search for new ferroelectric materials has ensued due to the presence of toxic lead in common $Pb[Zr_xTi_{1-x}]O_3$ (PZT). One promising lead-free candidate is $K_xNa_{1-x}NbO_3$. It was first shown in the 1950s that a solid solution of ferroelectric $KNbO_3$ and antiferroelectric $NaNbO_3$ induces a ferroelectric phase, which was then refined for piezoelectric applications approximately 50 years thereafter with an equal share of the constituents as $K_{0.5}Na_{0.5}NbO_3$ [19]. Several phases are now available in thin-film form with favorable ferroelectric properties due to their high Curie temperature of 690 K, polarization (33 $\mu C\,cm^{-2}$), and electromechanical planar coupling coefficient (0.454).

In the following, an imaging study of ferroelectric $K_{0.6}Na_{0.4}NbO_3$ (KNN) thin films on (110) $TbScO_3$ (TSO) substrates by means of scanning transmission x-ray microscopy (STXM) and holography-assisted CDI at the O *K*-edge is presented. The strongly anisotropic strain conditions of the TSO substrate lead to the formation of four different monoclinic superdomains consisting of periodic 90° ferroelectric stripe domain arrangements with the alignment of domain walls in the $[1\bar{1}2]_{TSO}$ or $[\bar{1}12]_{TSO}$ orthorhombic directions of the TSO substrate [8,20–22]. Charge neutrality at the domain walls and neighboring superdomains forces the ferroelectric in-plane polarization of the stripe domains to point along $[1\bar{1}0]_{TSO}$ or $[001]_{TSO}$. It follows that the effective ferroelectric polarization of each superdomain is oriented along the propagation direction of its ferroelectric stripe domains. The ferroelectric stripe period within individual superdomains need not be the same, as there are subtle differences in elastic boundary conditions that affect the interplay with the depolarization field and the energy associated with the domain walls.

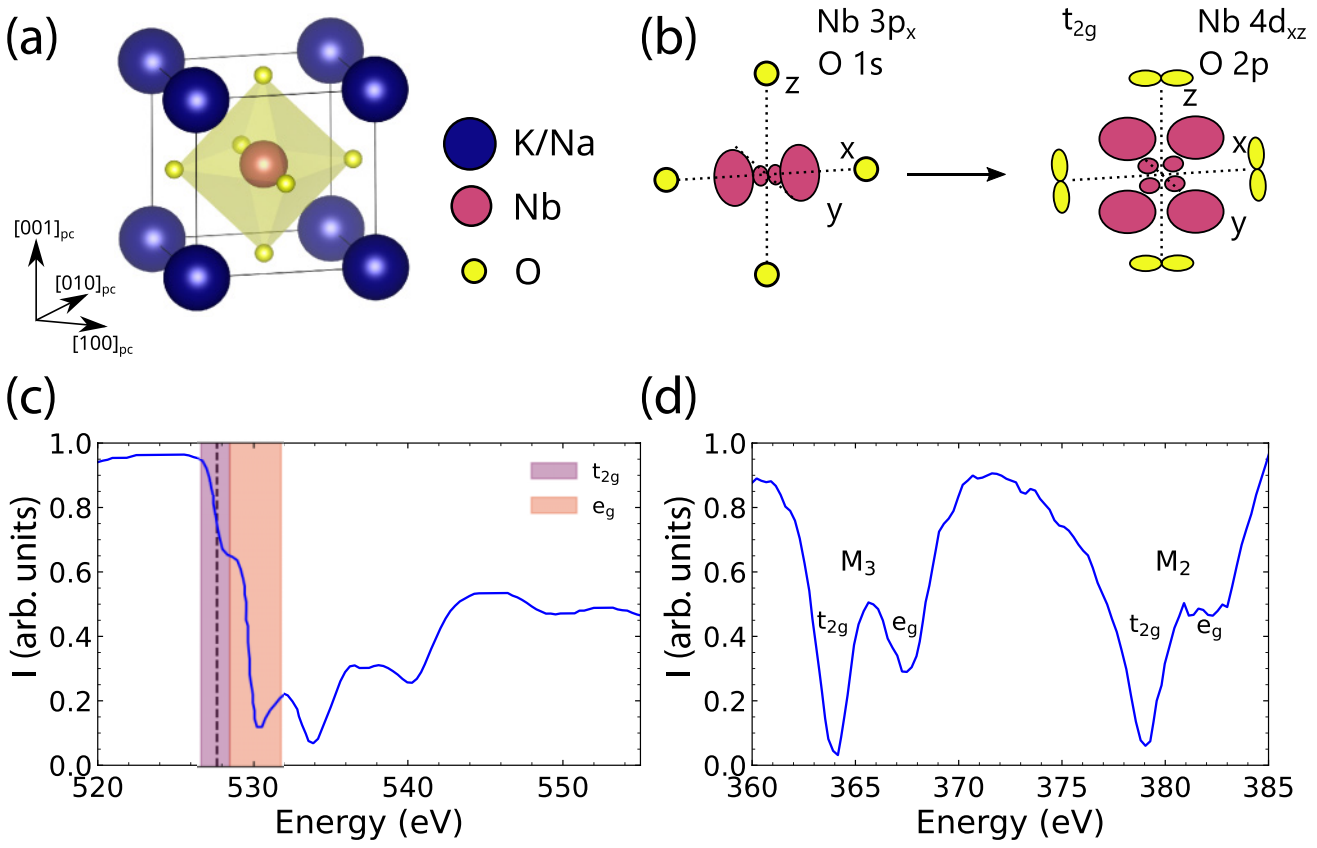

FIG. 1. (a) The pseudocubic (pc) perovskite unit cell of KNN shows the octahedral coordination of $Nb^{5+}$ by $O^{2-}$. (b) Sketch of the electronic orbitals in the ground (left) and excited hybridized state (right). The anisotropy in the O 2$p$ orbitals causes strong XLD at the O *K*-edge. (c) X-ray transmission spectrum of 37-nm KNN on TSO at the O *K*-edge. The peaks assigned to hybridization with 4$d$ $t_{2g}$ and $e_g$ are shaded. The broken vertical line indicates the photon energy of maximum XLD. (d) X-ray transmission spectrum of 37-nm KNN on TSO at the Nb $M_{3,2}$-edges.

The pseudocubic unit cell of KNN is displayed in Fig. 1(a) with the $O^{2-}$ octahedra surrounding the central Nb in the perovskite structure. The ferroelectricity of KNN largely originates in the off-center displacement of the pentavalent $Nb^{5+}$ cation inside the $O^{2-}$ octahedra [23]. It follows that the ferroelectric order is amenable to dichroic imaging at the O *K*-edge, which is sensitive to the O 2$p$–Nb 4$d$ hybridized excited states [24–31], analogously to ferroelectrics with 3$d$ transition metals such as BFO [5,13]. In octahedral symmetry, the 4$d$ states can point in direction of the $O^{2-}$ ions ($e_g$) or in between them ($t_{2g}$). The excitation process is shown schematically in Fig. 1(b) with the spherical O 1$s$ and the Nb 3$p$ states adjacent to one of the $t_{2g}$ hybridized state, which consists of the O 2$p$ and Nb 4$d_{xz}$ in the sketch. The anisotropic O 2$p$ orbitals cause the XLD by preferential absorption of x-rays with linear polarization parallel to the asymmetry in charge distribution. Under normal incidence of the x-rays, the in-plane projection of the ferroelectric polarization in the thin film can be sensed in this way and the XLD contrast inverts when the linear polarization is changed by 90°.

Two KNN thin films of 37- and 100-nm thicknesses were grown under compressive strain by liquid-delivery spin metal–organic chemical vapor deposition [32,33] on (110) $TbScO_3$ (TSO) substrates. The thin-film characterization is shown in the Supplemental Material [34]. The KNN sample of 100-nm thickness had an additional pulsed laser deposited 20-nm-thick $SrRuO_3$ (SRO) intermediate layer [22]. Both samples showed well-ordered stripe domains in PFM images that can be found in the Supplemental Material (Fig. S2 [34]). The superdomains contain stripes of around 50- and 100-nm period for the 37- and 100-nm films, respectively. Ferroelectric domains in KNN thin films scale with the square root of the film thickness according to Kittel's law [20]. In the case of the 100-nm-thick KNN film, the SRO interlayer is grown pseudomorphically on the TSO, which can be seen x-ray reciprocal space maps in Fig. S3. This means that the in-plane lattice parameter of the SRO is identical to that of KNN at the SRO/TSO and SRO/KNN interfaces. Thus, the KNN is grown fully strained to the TSO and there is no strain decoupling due to the SRO interlayer.

The strong absorption of soft x-rays presents a challenge for the imaging of epitaxial ferroelectric thin films on substrates that are far above the permissible thickness for transmission experiments. This is especially pertinent for x-ray microscopy of films on oxide substrates at the O $K$-edge. Hence, the TSO substrates were thinned mechanically to approximately 20 μm with a MultiPrep system (Allied High Tech Products). This approach is applicable to a wide range of common substrates, although the hardness of certain materials may make planar polishing of the entire sample impractical. Subsequently, square and circular membranes of approximately 25-μm diameter with TSO thicknesses below 1 μm for dichroic imaging were created by Ga-ion focused ion beam (FIB) for which a few nanometers of carbon were deposited to ensure conductivity and avoid charging [35,36]. The depth and surface morphology of the x-ray transparent windows was monitored *in situ* by conventional scanning electron microscope (SEM) at varying acceleration voltages. Examples of two adjacent square membranes in the TSO are shown in the inset of Fig. 2(a). In principle, it is possible to make larger x-ray transparent windows with the FIB, but these become prone to shattering at surface areas far below what can be achieved with $Si_3N_4$, which can range up to the millimeter range.

The x-ray absorption spectrum of the 37-nm-thin KNN sample on TSO was measured in transmission, both at the O $K$-edge ($1s \rightarrow 2p$) and the Nb $M_{3,2}$-edges ($3p_{3/2}/3p_{1/2} \rightarrow 4d$) with linear horizontal polarization. The resulting spectra are displayed in Figs. 1(c) and 1(d), respectively. The O $K$-edge spectrum shows the usual structure of a transition metal oxide around 529 eV with two crystal-field split absorption peaks due to hybridization, which are highlighted in Fig. 1(c). The O $K$-edge spectrum also contains contributions from the TSO substrate, which partially overlap with those of KNN. Both are perovskites with oxygen coordinating $Nb^{5+}$ or $Sc^{3+}$ octahedrally. The $4d$ states of Nb are further spatially extended than the $3d$ states of Sc, which causes greater hybridization and higher intensity features in the O $K$-edge spectrum. Conversely, Sc has only one $3d$ electron, the number of which is linked to the degree of hybridization and shift to lower energies in the O $K$-edge spectrum [29,37]. The distortion of the $NbO_6$ octahedra and the ferroelectricity of KNN is accessible by studying the hybridization with the Nb $4d$ states represented by the lowest-energy peaks of the spectrum. The Nb $4d$ states are also accessed at the Nb $M_{3,2}$-edges, which are shown in Fig. 1(d). The crystal field also splits these absorption peaks, which is significantly more perceptible than in previous spectroscopical studies of Nb oxides [26,30].

The ferroelectric domains in the x-ray transparent window for the 100-nm-thick KNN film on 20-nm SRO were imaged by STXM at the O $K$-edge with a FZP of 25-nm outer zone width and 240-μm diameter. The sample was aligned with the linear horizontal (LH) polarization of the incident x-rays from an APPLE II undulator parallel to the $[1\bar{1}0]_{TSO}$ direction for maximum XLD contrast. The result of an 18 μm × 18 μm STXM scan with horizontally polarized x-rays at the O $K$-edge is shown in Fig. 2(a). The photon energy corresponding to the $t_{2g}$ hybridization between the O $2p$ and Nb $4d$ states maximized the XLD contrast (dashed line in Fig. 1). The full x-ray absorption spectrum is shown in the Supplemental Material (Fig. S4 [34]). High XLD with the $t_{2g}$ hybridization has also been observed at the Ti $L_{2,3}$-edges in ferroelectric $PbTiO_3$ and PZT [38–40]. This is distinct from the behavior of the O $K$-edge in BFO in which the $e_g$ states show the maximum XLD contrast [13,41].

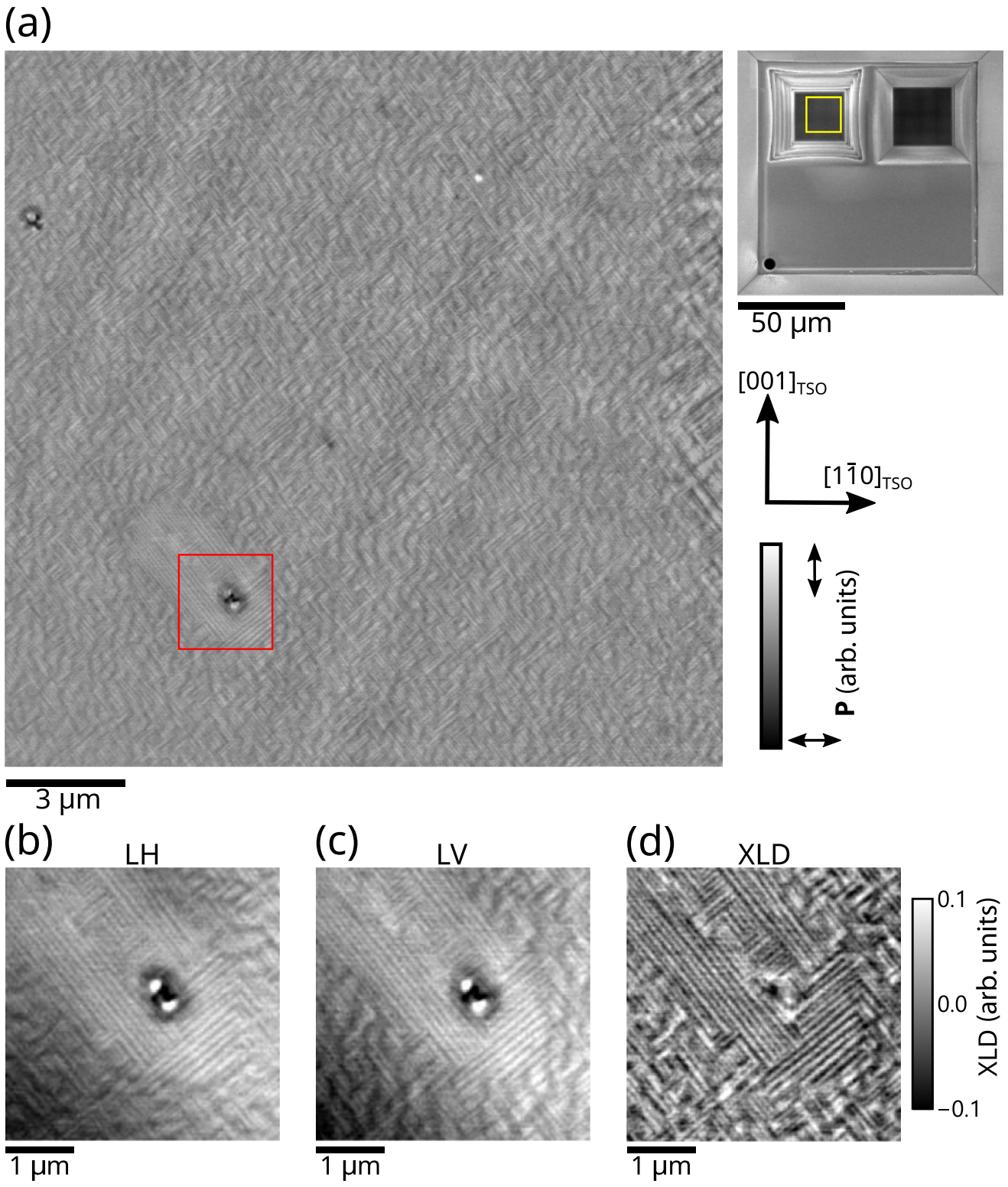

FIG. 2. (a) STXM image of ferroelectric domains in KNN in an 18 μm × 18 μm area. Horizontally polarized x-rays at the O $K$-edge tuned to the energy of $2p$-$4d\,t_{2g}$ hybridization were used. The in-plane components of the ferroelectric polarization **P** are in $[001]_{TSO}$ and $[1\bar{1}0]_{TSO}$ directions. An SEM image of two x-ray transparent windows in the substrate is shown in the top right with the imaged region framed in yellow. (b)–(d) Higher-resolution images of the framed area in (a) with LH/LV polarized x-rays and the corresponding XLD image, emphasizing the ferroelectric domains and suppressing topography contrast.

The XLD from the ferroelectric stripe domains is dominant in the STXM image and regions of different superdomains are discernible, which agrees well with the PFM images. The transmission image obtained with a single polarization also shows changes in contrast due to variations in the thickness of the TSO substrate and small circular defects in the film. The variations in the stripe domain periodicity and morphology originate in local decoupling of the 100-nm-thick KNN from the TSO substrate and SRO interlayer with strain gradients in growth direction [42,43]. While the stripe domain period of thicker films is higher, the size of the superdomains shrinks and a wider range of stripe domain periods is observed throughout the film that can relieve stress [20,22]. Furthermore, the metallic SRO interlayer may lead to screening of the surface charge density caused by the polarization of the KNN, which is significant for 20-nm SRO and may lead to the preference for certain directions of the ferroelectric polarization [44].

Nonferroelectric contributions to the image can be removed by subtraction of images recorded with LH and linear vertically (LV) polarized x-rays. This is demonstrated in Figs. 2(b)–2(d) for the area framed in red in Fig. 2(a). The contrast of the stripe domains with in-plane ferroelectric polarizations **P** along $[1\bar{1}0]_{\mathrm{TSO}}$ and $[001]_{\mathrm{TSO}}$ inverts between Figs. 2(b) and 2(c). The XLD image [Fig. 2(d)], calculated by $\mathrm{XLD} = \log(I_{\mathrm{LV}}/I_{\mathrm{LH}})$, thus separates ferroelectric from structural contributions, making the approach particularly valuable for epitaxial films on substrates with inherent thickness variations. Furthermore, the defects of higher transmittance in the center of the image are removed in the XLD contrast image and reveal their effect on the ferroelectric superdomains surrounding them. The mean stripe periods were 116 and 98 nm in the superdomains around the defect. The variation in stripe period and arrangement in vicinity of the defect suggest a local modification of epitaxial strain, illustrating how STXM can reveal nanoscale coupling between structural defects and ferroelectric order.

The periodic stripe ferroelectric texture of KNN, identified by PFM and confirmed by STXM, is an excellent candidate for an investigation by diffraction in reciprocal space: the mean stripe period, its distribution, and the angular alignment of the superdomains are encoded directly in the resonant diffraction pattern, averaged over the illuminated area and without the need for scanning. Moreover, the photon-energy dependence of the diffraction peaks isolates the electronic transitions that carry the dichroic ferroelectric contrast without any contribution from the substrate, resulting in an almost background-free detection of the ferroelectric signal. Resonant scattering from ferroelectric domains has so far been performed in reflection or Bragg geometry [39–41,45]. The substrate thinning demonstrated above lifts this restriction and allows the resonant scattering of an epitaxial ferroelectric film to be recorded in the forward direction, where the oversampled diffraction pattern can be converted into a real-space image by holography and phase retrieval approaches.

The spatial resolution of STXM is limited by the spot size created by the FZP, which is 1.2 times its outer zone width and is around 25 nm for regular operation. Unlike direct imaging methods such as STXM, the spatial resolution of CDI is only restricted by the detected photons at high scattering angles from which the phase can be recovered [46]. Limits for the resolution are set by the distance between sample and detector, the pixel size, the signal-to-noise ratio of the scattering signal, and the wavelength of the soft x-rays, which is approximately 2.3 nm at the O $K$-edge.

The thinner epitaxial film of 37-nm KNN with smaller domains was therefore studied by resonant x-ray scattering (RXS) and holography-assisted CDI [46] in transmission geometry. The electronic transitions probed by RXS are identical to those in x-ray absorption, which only probes the imaginary part of the complex atomic scattering factor responsible for absorption. Due to the sensitivity of RXS to the dispersive contributions to the resonant contrast, CDI techniques also provide phase contrast in a complex-valued image reconstruction.

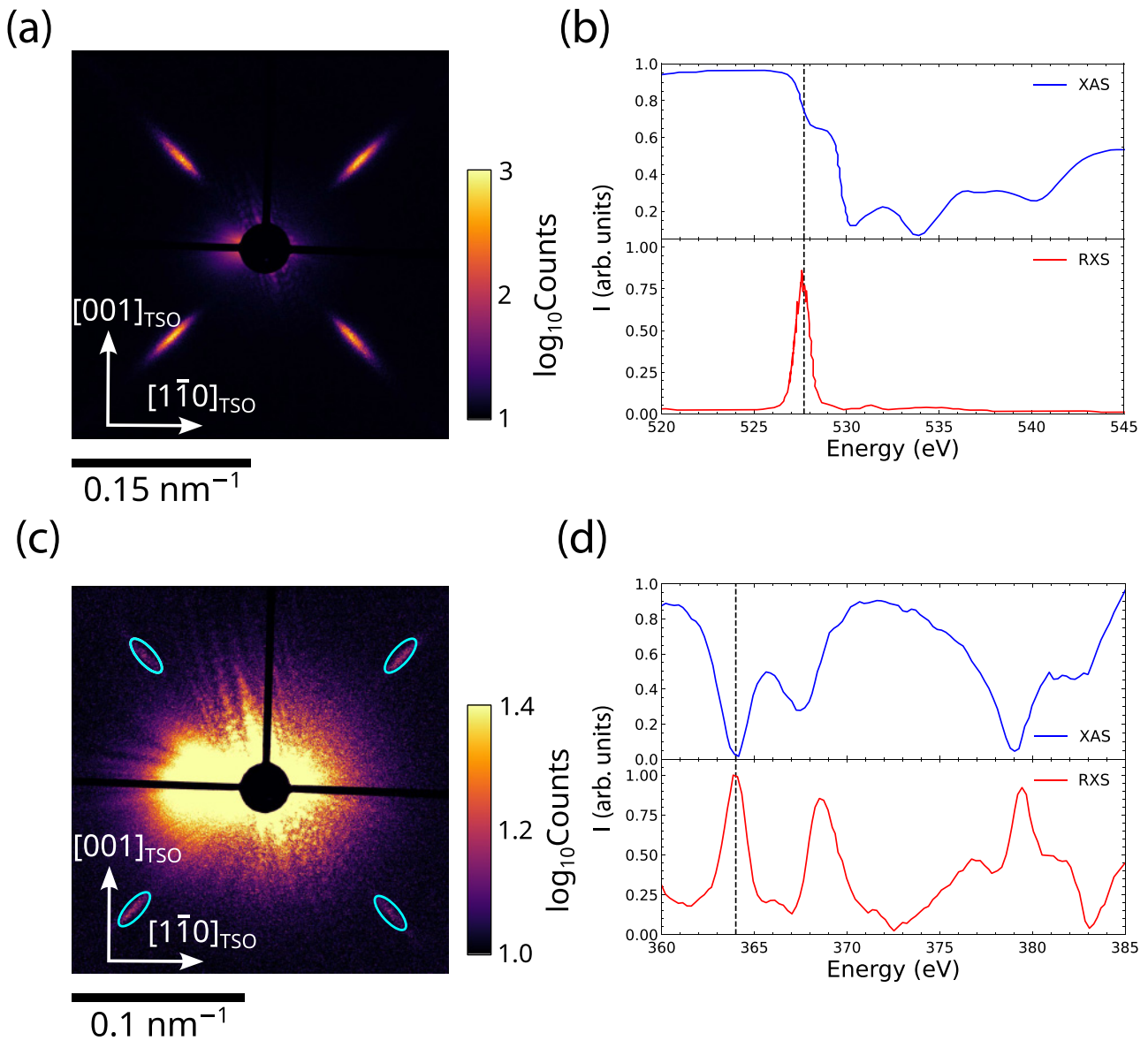

FIG. 3. Resonant x-ray scattering from ferroelectric domains in 37-nm KNN on (110) TSO. (a) RXS at the O $K$-edge (527.7 eV) shows the presence of diffraction peaks corresponding to the ferroelectric superdomains with ferroelectric stripes of 50-nm mean period in two directions. (b) The normalized x-ray transmission (XAS) and RXS signal from the diffraction peaks are shown. The maximum intensity of diffraction from the superdomains is at the $t_{2g}$ hybridization of the O $2p$ and Nb $4d$ states. (c) RXS signal at the Nb $M_3$-edge (364 eV) also shows the ferroelectric domains with lower intensity. (d) Intensity of the RXS signal is detectable at the crystal-field split transitions of the Nb $M_{3,2}$-edges.

The RXS diffraction patterns were recorded at the Veritas beamline of the MAX IV synchrotron-radiation source with a Princeton Instruments PI-MTE (MTE-2 1300B) CCD with 1340 × 1300 pixels of 20-μm size that was positioned 22 cm from the sample. A beamstop of 3.6-mm diameter on a cross was placed in front of the camera to block the intense zero-order x-ray beam. The photon energy was scanned in a range of 520–555 eV with LH polarization parallel to $[1\bar{1}0]_{\mathrm{TSO}}$ to identify the resonant interaction of x-rays and ferroelectric polarization at the O $K$-edge. The regularly spaced superdomains in the x-ray transparent window manifested themselves as peaks at the corresponding periodicity of their incorporated ferroelectric stripes in the diffraction pattern [see Fig. 3(a)]. The elongation of the diffraction peak indicates a variation of superdomain sizes and potentially nonuniformity of the stripe domains with a mean period of around 50 nm. Even so, the resonant diffraction peaks in the 37-nm thin KNN are significantly more well defined and pronounced than at the same photon energy in the 100-nm-thick sample with the SRO interlayer in which the KNN had epitaxially relaxed with the increased thickness (see Fig. S5 in the Supplemental Material [34]). The photon energy dependence of the x-ray transmission and the RXS signals are displayed in Fig. 3(b). The strongest resonant scattering intensity at 527.6 eV corresponds to the $t_{2g}$ hybridization between O $2p$ and Nb $4d$ states [see Fig. 1(c)], which also maximized XLD in STXM. The diffraction peaks of the ferroelectric superdomains are visible with 95% reduced intensity at 2.5-eV higher-photon energies. This transition can be assigned to the $e_g$ hybridization of O $2p$ and Nb $4d$ states [24,26].

Further RXS measurements were performed at the $M_3$ and $M_2$-edges of Nb between 360–385 eV. Although the RXS signal was diminished at lower photon energies, the intensities of the diffraction peaks due to the ferroelectric superdomains were still amplified at all the respective transitions [see Fig. 1(d)]. The RXS signal at the $M_3$ $t_{2g}$ transition with the lowest energy is displayed in Fig. 3(c), next to the x-ray transmission and integrated RXS signals in Fig. 3(d).

The diffraction peaks of the ferroelectric superdomains did not show dichroism for circularly polarized x-rays at either the O $K$-edge or the Nb $M_{3,2}$-edges. Thus, there is no clear indication of polar chirality in the stripe domain walls as was reported for 71° domains in BFO on scandate substrates [41,45] or in ferroelectric $PbTiO_3/SrTiO_3$ superlattices [39,40], which was confirmed by RXS in reflection geometry.

The conversion of the diffraction peaks pertaining to the ferroelectric domains at the O $K$-edge into a real-space image was accomplished by Fourier transform holography (FTH) [47] with phase retrieval [46], which requires a FTH mask on the sample. This consists of covering the membrane with a metallic Au/Cr multilayer mask, which blocks x-rays except in a circular aperture for the sample and smaller reference holes with diameters in the order of 100 nm. The circular window with which the RXS experiments were performed is shown in Fig. 4(a). The Au/Cr FTH mask was prepared separately, cut from the metallic film as a square and transferred to the KNN film side as shown in Fig. 4(b). A zoomed-in view of the FTH mask marked in Fig. 4(b) is shown in Fig. 4(c) after attachment to the KNN surface and Fig. 4(d) shows a view from the opposite side, i.e., through the remaining TSO substrate of sub-µm thickness. The field of view of the holographic image is defined by the diameter of the object hole, which was 1.2 µm for this sample. The imaging of extended areas such as by STXM [Fig. 2(a)] is impossible by FTH with a rigidly fixed mask. On the other hand, the spatial resolution cannot be adversely affected by uncertainties in sample positioning as is the case in STXM or ptychography.

A Princeton Instruments PI-MTE: 2048B CCD with 2048 × 2048 pixels of 13.5-µm size was placed 86 mm from the sample and a beamstop of 1-mm diameter was used for the FTH imaging. Illumination of the FTH sample by coherent x-rays at the photon energy providing maximum RXS contrast generated a coherent diffraction pattern, which constitutes an oversampled hologram with interference of x-rays scattered from the ferroelectric stripe domains and diffracted at the reference holes. An XLD map of the ferroelectric domain structure was reconstructed from the difference between diffraction patterns with LH and LV polarized x-rays [see Fig. 4(e)]. The result is shown in Fig. 4(f). The field of view includes three superdomains at 90° with stripe periods of 57 and 44 nm, which are well ordered due to the epitaxial strain from the TSO. The clearly resolved stripe down to 44-nm period demonstrates that holography-assisted CDI substantially surpasses the resolution limit of STXM. Estimation of the spatial resolution with ferroelectric sensitivity is possible by Fourier ring correlation, which is shown in the Supplemental Material (Fig. S6 [34]) and indicates 18.6 nm. This implies the highest reliably detected frequency for this particular sample and the value may be exceeded in samples with smaller ferroelectric domains. The FTH geometry employed here is directly compatible with emerging real-time imaging schemes such as coherent correlation imaging [48] and with ultrafast pump-probe approaches for reproducible processes at x-ray free-electron laser (XFEL) [49,50] or high-harmonic generation sources [51], opening prospects for femtosecond imaging of ferroelectric domain dynamics.

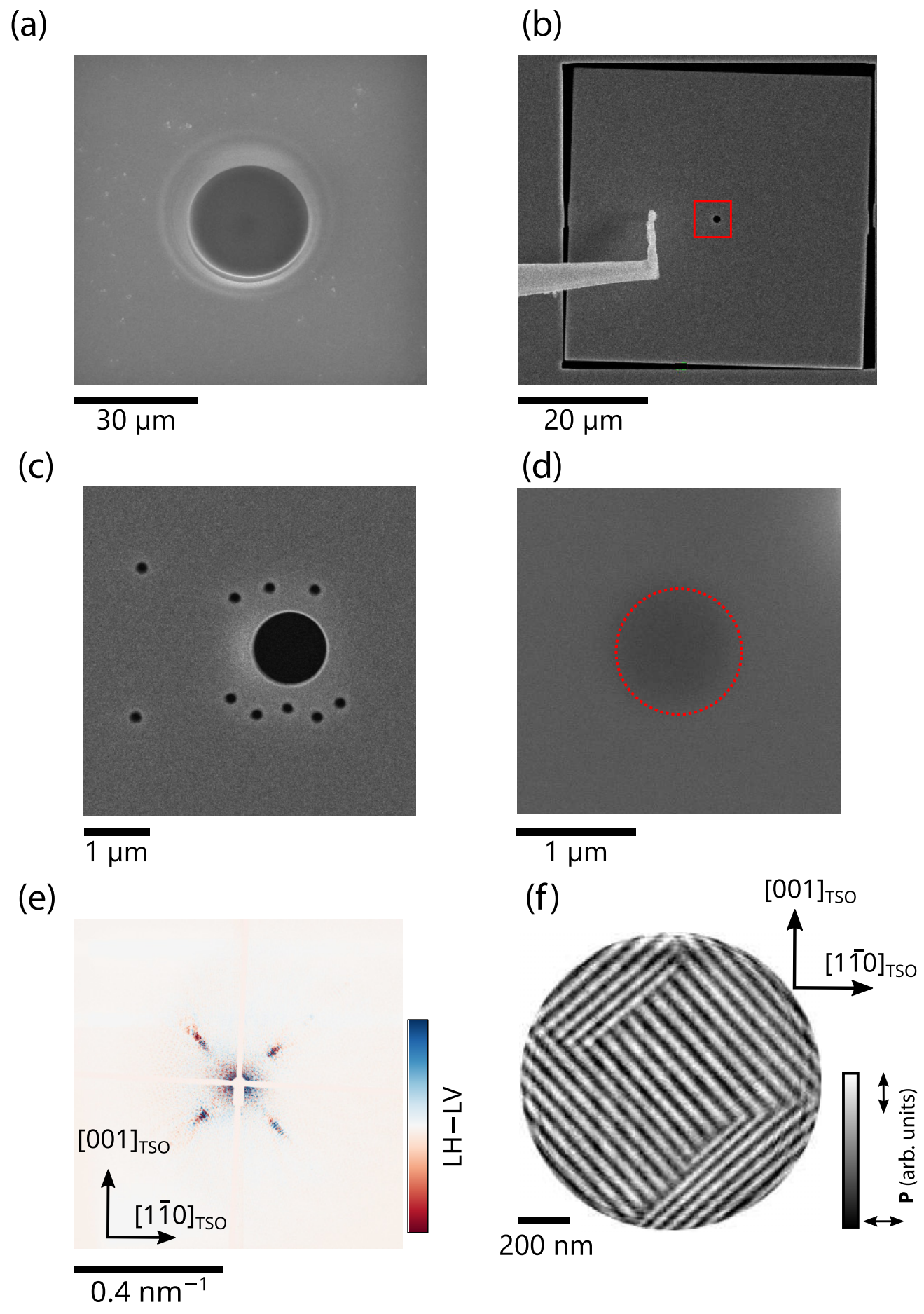

FIG. 4. Holography-assisted CDI of ferroelectric domains in KNN on TSO. (a) X-ray transparent window for the RXS measurements in Fig. 3. (b) Square Au/Cr multilayer FTH mask during transfer to the KNN sample. (c) Enlarged view of the FTH mask with object and reference holes after fastening to the sample surface, corresponding to the red framed area in (b). (d) FTH sample from the KNN side. The transparent circular window for the object is circled. (e) Difference in the diffraction patterns with LH and LV polarization obtained at the maximum ferroelectric RXS at the O $K$-edge [see Figs. 3(a) and 3(b)]. The superdomain diffraction peaks appear with interference fringes. (f) Reconstructed holographic XLD difference image of the ferroelectric superdomains with 57-nm and 44-nm in-plane stripe periods.

In conclusion, the ferroelectric domain structure of epitaxially strained KNN thin films on (110) TSO was imaged at the nanoscale by linear dichroic imaging with STXM and holography-assisted phase retrieval. The ferroelectric stripe domains of the compressively strained KNN films remained intact after back-thinning of the substrate to below 1 μm for soft x-ray microscopy in transmission. The O *K*-edge is sensitive to ferroelectric order in KNN due to the hybridization with Nb 4*d* states, causing strong XLD and RXS signals for the in-plane components of the ferroelectric polarization in $[001]_{\mathrm{TSO}}$ and $[1\bar{1}0]_{\mathrm{TSO}}$ directions.

The ferroelectric polarization in KNN also has an out-of-plane component, which is not accessible by XLD at normal incidence. Characterizing the three-dimensional orientation of the polarization axis requires tilting the sample with respect to the incident x-rays. In the case of thin films, laminography is better suited to extended planar samples than the full rotation used for linear dichroic tomography with hard x-rays [52] because tilting the rotation axis keeps the sample thickness constant and avoids excessive absorption at high angles.

The spatial resolution of STXM was improved upon by reconstruction of the superdomain diffraction peaks from RXS by holography-assisted CDI. The substrate thinning approach demonstrated here is also directly applicable to ptychographic imaging [12,13,16,17,53–55]. In future experiments, the spatial resolution and sensitivity of these coherent-diffraction-based imaging approaches will be further enhanced by the availability of single-photon counting area detectors with low readout noise and higher dynamic range in the soft x-ray regime [56]. These improvements will allow detection of faint scattering from ferroelectric domains of metallic oxides with weaker hybridizations. The successful x-ray imaging of ferroelectric order opens possibilities for time-resolved imaging of ferroelectric dynamics, surpassing what has been achieved by stroboscopic PFM with temporal resolutions limited to 100 ns [57]. Using STXM, 50-ps temporal resolutions can be obtained at synchrotrons with avalanche photodiodes [58–60] and are already routinely used for the study of magnetization dynamics [53]. Higher-resolution CDI imaging methods will be valuable for the investigation of switching and laser-induced dynamics [9,61] at XFEL facilities [50].

We thank the European Regional Development Fund (ERDF) for funding (Project No. 1.8/15) and S. Bin Anooz for film growth. The $TbScO_3$ substrates were grown at the Leibniz-Institut für Kristallzüchtung in the group of S. Ganschow. STXM measurements were performed at the MAXYMUS and MYSTIIC end stations at the BESSY II electron storage ring operated by the Helmholtz-Zentrum Berlin (HZB) für Materialien und Energie. We thank HZB for the allocation of synchrotron-radiation beamtime. FTH assisted CDI measurements were performed with the MAXI endstation of the Max Born Institute at the Veritas and SoftiMAX beamlines of the MAX IV Laboratory. Research conducted at MAX IV, a Swedish national user facility, is supported by the Swedish Research council under Contract No. 2018-07152, the Swedish Governmental Agency for Innovation Systems under Contract No. 2018-04969, and Formas under Contract No. 2019-02496. We acknowledge DESY (Hamburg, Germany), a member of the Helmholtz Association HGF, for the provision of experimental facilities. Supporting measurements for this research were carried out at PETRA III and we would like to thank S. Chowdhury and M. Hoesch for assistance in using beamline P04. T.A.B. acknowledges funding from the ERDF and the Röntgen–Ångström Cluster grant "X-MOOSE" (Grant No. BMFTR 05K2026). We acknowledge funding from the Helmholtz Young Investigator Group Program through Project No. VH-NG-1520, from the Deutsche Forschungsgemeinschaft (DFG, German Research Foundation) through Projects No. 462676630 (BiSky), No. 505818345 (Topo3D), and No. 49254781 (TRR 360), and the German Federal Ministry of Education and Research (BMBF) under Grant No. 05K24BCA (Soft-XPCS) within the Röntgen–Ångström Cluster. We thank M. Brede for assistance during the STXM measurements.

*Data availability*. The data that support the findings of this article are openly available [62].